\documentclass[aps,twocolumn,prl,showpacs]{revtex4-1}
\usepackage{bm}
\usepackage{amsmath}
\usepackage{amssymb}
\usepackage{graphicx}

\begin{document}

\title{Spin responses and effective Hamiltonian for the two dimensional
electron gas at oxide interface {LaAlO}$_3$/{SrTiO}$_3$}

\author{Jianhui Zhou}
\email{jhzhou@andrew.cmu.edu}
\affiliation{Department of Physics, Carnegie Mellon University, Pittsburgh, Pennsylvania 15213, USA}

\author{Wen-Yu Shan}
\email{wyshan@andrew.cmu.edu}
\affiliation{Department of Physics, Carnegie Mellon University, Pittsburgh, Pennsylvania 15213, USA}

\author{Di Xiao}
\affiliation{Department of Physics, Carnegie Mellon University, Pittsburgh, Pennsylvania 15213, USA}

\begin{abstract}
Strong Rashba spin-orbit coupling (SOC) of two-dimensional electron
gas (2DEG) at the oxide interface $\mathrm{LaAlO_{3}/SrTiO_{3}}$ underlies 
a variety of exotic physics, but its nature is still under debate.
We derive an effective Hamiltonian for the
2DEG at the oxide interface $\mathrm{LaAlO_{3}/SrTiO_{3}}$ and find
a different anisotropic Rashba SOC for the $d_{xz}$ and $d_{yz}$ orbitals.
This anisotropic Rashba SOC leads to anisotropic static spin susceptibilities and also distinctive behavior of the spin Hall conductivity.  These unique spin responses
may be used to determine the nature of the Rashba SOC experimentally and shed light on the orbital origin of the 2DEG.
\end{abstract}


\maketitle


The discovery of a high mobility two-dimensional electron
gas (2DEG) at the interface between two band insulators $\mathrm{LaAlO_{3}}$
and $\mathrm{SrTiO_{3}}$ (LAO/STO) \cite{OhtomoHwang} has attracted
increasing attention \cite{HwangReview}. However, the origin of the 2DEG is still under active debate. According to the intrinsic polar catastrophe mechanism, there should be a half electron (per unit cell) transfer from the top surface layer of LAO to the LAO/STO interface. The resulting
carrier density at the interface is roughly $3.5\times10^{14}$~cm$^{-2}$,
which mainly comes from the three $t_{2g}$ orbitals of $\mathrm{Ti}$
in STO.  Several transport experiments, however, estimate that
the carrier density is only $10\%$ of that due to the polar catastrophe mechanism
\cite{CavigliaNature,Bell,ThielScience}.  In addition, it has been proposed that electrons in the $d_{xy}$ orbitals, which are confined in the $xy$~plane, are more likely to become localized at the
interface due to the impurities or electron-phonon coupling, while
those in the $d_{xz}$ and $d_{yz}$ orbitals are itinerant
and contribute to transport~\cite{Popovic09PRL}. Within this scenario,
the localized and itinerant electrons would account for the observed magnetic
order \cite{JSLeedxyFM} and superconductivity \cite{ReyrenSC,BertFMSC,DikinFMSC,LuLiNatPhysFMSC},
respectively.  It is therefore important to understand the transport properties of the 2DEG.  Other mechanisms, such as oxygen vacancies~\cite{KalabukhovOV,HerranzOV} and polar distortion~\cite{HamannPD,SatoshiPD}, have also been proposed.

Recent magnetotransport experiments have provided us insight into
the 2DEG at the oxide interface. In particular, 
a strong and field-tunable Rashba spin-orbit coupling (SOC) was observed \cite{ShalomPRL10SOC,CavigliaPRL10SOC} and was modeled using the standard $k$-linear form \cite{BychkovRashba}, i.e.,
\begin{equation}
H_R = \lambda_R (\bm k \times \boldsymbol{\sigma}) \cdot \hat{z} \;.
\end{equation}
Based on this $k$-linear Rashba SOC, theoretical
works have predicted a variety of unusual effects, such as Fulde-Ferrell-Larkin-Ovchinikov-type
superconductivity coexisting with ferromagnetism \cite{Michaeli},
spiral magnetic order and skyrmions \cite{Banerjee,XPLi,BanerjeePRX,Scheurer}, and  
the spin Hall effect \cite{HaydenSHE}. However, a very recent magneto-conductivity
measurement has suggested the possibility of a $k$-cubed Rashba SOC of the 2DEG
at the oxide interface \cite{Nakamura,KimNayak}. Accordingly, some authors proposed a 
$k$-linear Rashba SOC for the $d_{xy}$ orbital \cite{Khalsa,ZCZhong}
and a $k$-cubed one for the $d_{xz}$ and $d_{yz}$ orbitals
\cite{ZCZhong}.  On the other hand, first-principles calculations
combined with the envelope function method have found an
anisotropic nonparabolic spin-split subband structure for the $d_{xz}$ and $d_{yz}$ orbitals \cite{HeeringenKP}, which
could not be explained by the standard $k$-cubed Rashba SOC.
Thus, a detailed investigation of the low energy effective model and the
nature of the Rashba SOC is highly desirable.

In this Rapid Communication, we present a detailed derivation of the effective Hamiltonian of the 2DEG
at the oxide interface.  We find a different anisotropic Rashba SOC of the following form,
\begin{equation}
H^\text{ani}_R \propto \left(k_{x}^{2}-k_{y}^{2}\right)\left(\boldsymbol{k}\times \boldsymbol{\sigma}\right)\cdot\hat{z},
\end{equation}
 for the $ d_{xz}$ and $d_{yz}$ orbitals, and a standard $k$-linear Rashba SOC for the $d_{xy}$ orbital.
The anisotropy of the Rashba SOC naturally leads to anisotropic spin susceptibilities that have been observed experimentally \cite{LuLiNatPhysFMSC,BertFMSC}.
We also show that this anisotropic Rashba SOC results in different behavior of the spin Hall conductivity (SHC) when compared to the standard $k$-linear and $k$-cubed Rashba SOCs.
These distinctive spin responses can be used for determining the nature of the Rashba SOC in experiments and to shed light on the orbital origin of the 2DEG at the LAO/STO interface.

%


We begin by constructing the low-energy effective model of the 2DEG
at the LAO/STO interface around the $\Gamma$ point in the Brillouin
zone. The 2DEG is formed from the $d$~orbitals of the transition-metal
Ti.  Here we focus on the three $t_{2g}$ orbitals, namely, $d_{xy}$, $d_{xz}$, and $d_{yz}$, since the $e_{g}$ orbitals are pushed up about 2~eV higher
than the $t_{2g}$ orbitals by the octahedral crystal field.
On the $xy$~plane, electrons in the $d_{xy}$ orbital can hop along either the $x$ or $y$ direction to the $d_{xy}$ orbitals
on the neighboring $\mathrm{Ti}$, while electrons in the $d_{xz}\left(d_{yz}\right)$ orbital can
hop to its neighbor only along the $x\left(y\right)$ direction.  Thus, the corresponding hopping 
Hamiltonian can be expressed in the following matrix form,

%
\begin{alignat}{1}
H_{0} & =\left(\begin{array}{ccc}
h(\boldsymbol{k}) & 0 & 0\\
0 & -2t\cos k_{x} & 0\\
0 & 0 & -2t\cos k_{y}
\end{array}\right),
\end{alignat}
where $h(\boldsymbol{k})=-\Delta_{E}-2t\left(\cos k_{x}+\cos k_{y}\right)$, $t=t_{pd}^{2}/\Delta_{pd}$ is the effective hopping parameter
between nearest neighboring $\mathrm{Ti}$, $\Delta_{pd}$ is the
splitting between the oxygen $p$ and $\mathrm{Ti}$ $t_{2g}$ energy
levels, and $\Delta_{E}$ is the difference in the on-site energies
between the $d_{xy}$ orbital and the $d_{yz}$/$d_{xz}$ orbital.  Note that since $d_{xy}$ is even, and $d_{xz}$ and $d_{yz}$ are odd under the operation $z\to -z$, hopping between these two sets of orbitals is prohibited in the presence of the mirror symmetry.

To model the effect of the SOC, we introduce the atomic SOC $H_{\xi}=\xi \boldsymbol{l}\cdot \boldsymbol{\sigma}$ in the basis $\left\{ \left|d_{xy}\uparrow\right\rangle ,\left|d_{xy}\downarrow\right\rangle ,\left|d_{xz}\uparrow\right\rangle ,\left|d_{xz}\downarrow\right\rangle ,\left|d_{yz}\uparrow\right\rangle ,\left|d_{yz}\downarrow\right\rangle \right\} $,
\begin{alignat}{1}
H_{\xi} & =\xi\left(\begin{array}{cccccc}
0 & 0 & 0 & -i & 0 & 1\\
0 & 0 & -i & 0 & -1 & 0\\
0 & i & 0 & 0 & -i & 0\\
i & 0 & 0 & 0 & 0 & i\\
0 & -1 & i & 0 & 0 & 0\\
1 & 0 & 0 & -i & 0 & 0
\end{array}\right),
\end{alignat}
where $\xi$ denotes the strength of the atomic SOC. $\boldsymbol{\sigma}$
refers to the spin degree of freedom, while $\boldsymbol{l}$
is the orbital angular momentum of the electron.

Finally, there is a mirror symmetry breaking at the interface due to the polar displacement of Sr and Ti atoms
relative to the oxygen octahedra, which leads to the Rashba SOC.
Physically, the mirror symmetry breaking can induce the hopping process from the $d_{xz}\left(d_{yz}\right)$ orbital
to the $d_{xy}$ orbital via the $p_{x}\left(p_{y}\right)$ orbital of oxygen. The corresponding Hamiltonian 
can be written as \cite{Khalsa,ZCZhong}
\begin{alignat}{1}
H_{\gamma} & =\gamma\left(\begin{array}{ccc}
0 & -2i\sin k_{y} & -2i\sin k_{x}\\
2i\sin k_{y} & 0 & 0\\
2i\sin k_{x} & 0 & 0
\end{array}\right)\otimes\sigma_{0},
\end{alignat}
where $\gamma$ refers to the effective hopping amplitude between
the $d_{xy}$ orbital and the $d_{xz}$ and $d_{yz}$ orbitals. $\sigma_{0}$ is the
$2\times2$ unit matrix in the real spin space. 

The total tight-binding (TB) Hamiltonian including all three parts is given by
\begin{alignat*}{1}
H_{\mathrm{TB}} & =H_{0}+H_{\xi}+H_{\gamma} \;.
\end{alignat*}
There are three pairs of degenerate bands at the $\Gamma$ point, which are plotted in Fig.~1(a) using the parameters given in Ref.~\onlinecite{ZCZhong}. 
It can be seen that the energy contour of the middle two bands has a strong anisotropy
as shown in Fig.~1(b), whereas the lowest two bands are isotropic as shown in Fig.~1(c).
Note that the splitting of the two lowest-energy bands due to the Rashba SOC is unnoticeable for the given energy.

To derive the effective Hamiltonian, we apply the quasidegenerate perturbation theory \cite{Winkler}.  Up to leading order in the
 SOC strength $\xi$, we obtain the effective Hamiltonian for the top pair of bands,
\begin{alignat}{1}
H_{\mathrm{top}}\left(\boldsymbol{k}\right) & =\frac{k^{2}}{2m_{\mathrm{top}}}-\alpha_{\mathrm{top}}\left(\boldsymbol{k}\times\boldsymbol{\sigma}\right)\cdot\hat{z},
\end{alignat}
the middle pair of bands,
\begin{alignat}{1} \label{OK}
H_{\mathrm{mid}}\left(\boldsymbol{k}\right) & =\frac{k^{2}}{2m_{\mathrm{mid}}}+\alpha_{\mathrm{mid}}\left(k_{x}^{2}-k_{y}^{2}\right)\left(\boldsymbol{k}\times\boldsymbol{\sigma}\right)\cdot\hat{z},
\end{alignat}
and the bottom pair of bands,
\begin{alignat}{1}
H_{\mathrm{bot}}\left(\boldsymbol{k}\right) & =\frac{k^{2}}{2m_{\mathrm{bot}}}-\alpha_{\mathrm{bot}}\left(\boldsymbol{k}\times\boldsymbol{\sigma}\right)\cdot\hat{z},
\end{alignat}
where $\left(m_{\mathrm{top}},m_{\mathrm{mid}},m_{\mathrm{bot}}\right)$
and $\left(\alpha_{\mathrm{top}},\alpha_{\mathrm{mid}},\alpha_{\mathrm{bot}}\right)$ are the effective
masses and Rashba SOC strengths for the top, middle, and bottom pairs
of bands, respectively (all their specific expressions are given in the Supplemental Material~\cite{SM1}). The top pair of bands is a mixture of all three $t_{2g}$
orbitals. The bottom pair mainly comes from the $d_{xy}$ orbital. The
middle pair is a hybridization between the $d_{xz}$ orbital and the $d_{yz}$ orbital.
It is also clear that the bottom pair of bands has the $k$-linear Rashba SOC
 that was proposed by previous works~\cite{Khalsa,ZCZhong}. 
 This concentric isotropic Fermi contour of the $d_{xy}$ orbital had also been demonstrated at the surface of bare $\mathrm{SrTiO_{3}}$ \cite{Meevasana}.
 In the middle pair of bands, the Rashba 
SOC becomes anisotropic and
has a $k$-cubed energy dispersion~\cite{touching}.
Two recent angle-resolved photoemission experiments had already observed the anisotropic 
Fermi contour of the $d_{xz}$ orbital and the $d_{yz}$ orbital at a high carrier density \cite{Cancellieri,PDKing}.  Note that the effective Hamiltonian of each pair of bands is constructed with respect to its own bottom edge.

%
\begin{figure}
\includegraphics[scale=0.44]{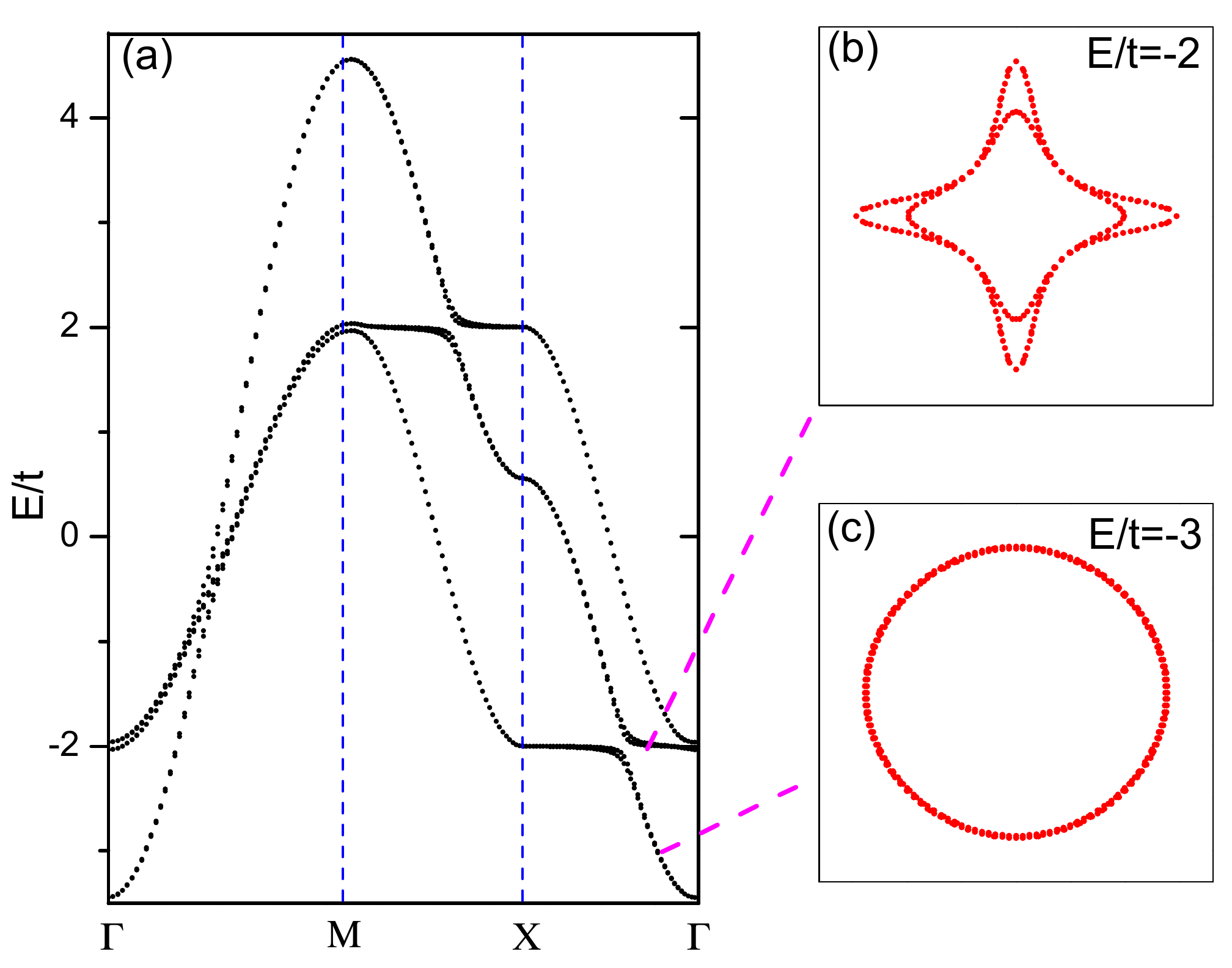}
\protect\caption{(Color online) 
(a) Band structure of TB model describing the oxide interface. Energy contours near the $\Gamma$ point for energies 
(b) $E/t=-2$ and (c) $E/t=-3$. Parameters are adopted from Ref. \onlinecite{ZCZhong}: $\Delta_E/t=-0.56$, $\xi/t=0.035$, $\gamma/t=0.072$.}
\label{fig:dispersion}
\end{figure}
%

%
%
%

The anisotropic Rashba SOC for the $d_{xz}$ and $d_{yz}$ orbitals in Eq.~\eqref{OK} is our main result.
In the rest of this Rapid Communication, we will study its effects on
 the  static spin susceptibility and the spin Hall conductivity~\cite{MurakamiScience,SinovaPRL}.  
For convenience, we redefine the corresponding effective mass $m=m_{\mathrm{mid}}/\hbar^{2}$
and Rashba SOC strength $\beta=\alpha_{\mathrm{mid}}/\hbar^{3}$. 
The effective Hamiltonian for the middle pair can be recast into
\begin{equation}
H_{\mathrm{mid}}\left(\boldsymbol{k}\right)=\left(\begin{array}{cc}
\frac{\hbar^{2}k^{2}}{2m} & -i\beta\hbar^{3}\left(k_{x}^{2}-k_{y}^{2}\right)k_{-}\\
i\beta\hbar^{3}\left(k_{x}^{2}-k_{y}^{2}\right)k_{+} & \frac{\hbar^{2}k^{2}}{2m}
\end{array}\right),
\end{equation}
with $k_{\pm}=k_{x}\pm i k_{y}$.
Some simple algebra leads to the eigenvalues of $H_{\mathrm{mid}}\left(\boldsymbol{k}\right)$,
\begin{equation}
\varepsilon_{\boldsymbol{k}s}=\frac{\hbar^{2}k^{2}}{2m}+s\beta\hbar^{3}k^{3}\left|\cos2\theta_{\boldsymbol{k}}\right|,
\end{equation}
and the corresponding eigenvectors,
\begin{equation}
\phi_{\boldsymbol{\boldsymbol{k}}s}=\frac{1}{L}e^{i\boldsymbol{k}\cdot\boldsymbol{r}}\eta_{\boldsymbol{k}s},
\end{equation}
where the spinor is given by $\eta_{\boldsymbol{k}s}=\left(-is\varsigma_{\boldsymbol{k}}e^{-i\theta_{\boldsymbol{k}}},1\right)^{T}/\sqrt{2}$,
$s=\pm1$ is the chirality index, and $L^{2}$ is the area of the 2DEG with 
$\varsigma_{\boldsymbol{k}}=\cos2\theta_{\boldsymbol{k}}/\left|\cos2\theta_{\boldsymbol{k}}\right|=\pm1$, $\theta_{\boldsymbol{k}}=\arctan(k_y/k_x)$.
Since our model is only valid around the $\Gamma$ point,  
we would like to introduce a momentum cutoff $k_{c}=1/3m\hbar\beta$ 
via the turning point of the energy dispersion
$\varepsilon_{k}=k^{2}\hbar^{2}/2m-\beta\hbar^{3}k^{3}$. The corresponding
energy of this turning point is given by $\varepsilon_{\mathrm{turn}}=1/54m^{3}\beta^{2}$.
 
 In general, the free spin susceptibilities can be written as
\begin{equation}
\chi_{ij}\left(\boldsymbol{q}\right)=-k_{B}T \mu_{B}^{2}\sum_{n,\boldsymbol{k}}\mathrm{Tr}\left[\sigma_{i}G\left(\boldsymbol{k},\omega_{n}\right)\sigma_{j}G\left(\boldsymbol{k}+\boldsymbol{q},\omega_{n}\right)\right],
\end{equation}
where $\sigma_{i}$ are the Pauli matrices with $i=x,y,z$, $G\left(\boldsymbol{k},\omega_{n}\right)$
is the Matsubara Green's function of an electron with momentum $\boldsymbol{k}$
and frequency $\omega_{n}$, and $\mu_{B}$ is the Bohr magneton. 
After carrying out the standard analytic continuation and frequency summation 
(more details of the derivation can be found in the Appendix of Ref.~\onlinecite{Zak}),
we can find the static spin susceptibilities in the limit $\boldsymbol{q}\rightarrow \boldsymbol{0}$,
\begin{align}
\chi_{zz}^{0} & =-2\mu_{B}^{2}\sum_{\boldsymbol{k}}\frac{f\left(\xi_{+}\left(\boldsymbol{k}\right)\right)-f\left(\xi_{-}\left(\boldsymbol{k}\right)\right)}{\xi_{+}\left(\boldsymbol{k}\right)-\xi_{-}\left(\boldsymbol{k}\right)},\\
\chi_{xx}^{0} & =-\frac{\mu_{B}^{2}}{2}\sum_{\boldsymbol{k},\lambda}\frac{\partial f\left(\xi_{\lambda}\left(\boldsymbol{k}\right)\right)}{\partial\xi_{\lambda}\left(\boldsymbol{k}\right)}+\frac{\chi_{zz}^{0}}{2},\\
\chi_{yy}^{0} & =\chi_{xx}^{0},
\end{align}
where $\xi_{\lambda}\left(\boldsymbol{k}\right)=\varepsilon_{\lambda}\left(\boldsymbol{k}\right)-E_{F}$
is the energy of the electron measured relative to the Fermi energy $E_{F}$, and
the superscript $0$ indicates the spin susceptibility with $\boldsymbol{q}=\boldsymbol{0}$.
All of the other components vanish due to the symmetry of Fermi
surface. The out-of-plane component $\chi_{zz}^{0}$ is the so-called
van Vleck susceptibility and originates from the virtual inter band
transition. The in-plane component $ $$\chi_{xx}^{0}$ or $\chi_{yy}^{0}$
contains both the intraband contribution (the Pauli susceptibility) and the interband contribution (the van Vleck susceptibility). 
Numerical calculations of the spin susceptibilities of 2DEGs with the anisotropic
Rashba SOC show two main features, as shown in Fig.~2. First, the spin susceptibilities
are anisotropic, i.e., $ $ $\chi_{zz}^{0}\neq\chi_{xx}^{0}$. Secondly,
the spin susceptibilities have a strong Fermi energy dependence. Note that the momentum cutoff $k_{c}$
is used in our numerical calculations.

Previously the anisotropic spin susceptibility was also found using the $k$-linear Rashba model~\cite{Fischer}.  However, the spin susceptibility is anisotropic when only the lower Rashba spin-split band is occupied.  As soon as both spin-split bands are occupied, the spin susceptibility becomes isotropic~\cite{Zak}.  As such, the anisotropy only shows up in a small energy window.  In contrast, the spin susceptibility in our model is always anisotropic (up to the turning point when the model is no longer valid)~\cite{SUSCR}.  Therefore, our result may provide an alternative explanation for the observed magnetic anisotropy~\cite{LuLiNatPhysFMSC,BertFMSC}.
%
\begin{figure}[t]
\includegraphics[scale=0.35]{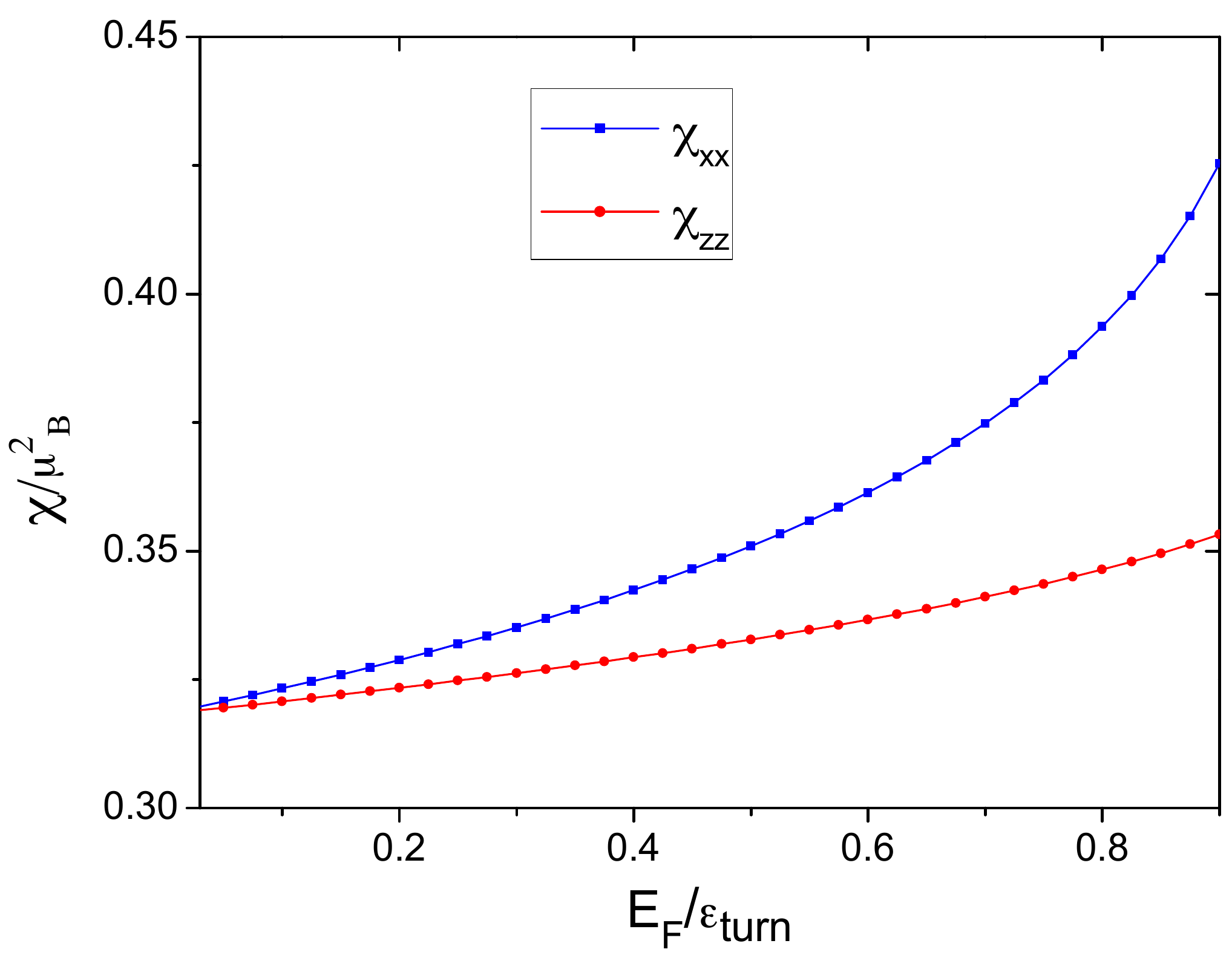}
\protect\caption{(Color online) The spin susceptibility of 2DEG with the anisotropic Rashba
SOC as a function of the Fermi energy $E_{F}$ in units of $\varepsilon_{\mathrm{turn}}$ (measured from the bottom of the middle pair of bands). 
We set the dimensionless effective mass of the electron $m=1$, the dimensionless Rashba SOC parameter ${\beta}=0.01$, and the temperature $T$=5~K.}
\end{figure}
 %
 
 
%

Let us now turn to calculate the SHC of the 2DEG with the anisotropic Rashba SOC. 
The general spin conductivity tensor in the spin space is given as
\begin{align}
\sigma_{\alpha x}^{\sigma_{i}} & =\frac{\hbar}{2\pi L^{2}}\sum_{\boldsymbol{k}}\mathrm{Tr}\left[J_{\alpha}^{\sigma_{i}}\tilde{K}_{x}\right],\label{SHC}
\end{align}
where $\tilde{K}_{x}\equiv UK_{x}U^{\dagger}$ is the vertex function
in the spin space and $K_{x}=\tilde{G}^{R}J_{x}\tilde{G}^{A}$ is the
vertex function in the eigenvectors space of $H_{\mathrm{mid}}(\boldsymbol{k})$.
$ $$\tilde{G}^{R}$ and $\tilde{G}^{A}$ are the retarded and advanced Green's function of 2DEG,
\begin{equation}
\tilde{G}_{\boldsymbol{k}s}^{A}\left(\epsilon\right)=\frac{1}{\epsilon-\varepsilon_{\boldsymbol{k}s}-i\eta},\:\tilde{G}_{\boldsymbol{k}s}^{R}\left(\epsilon\right)=\frac{1}{\epsilon-\varepsilon_{\boldsymbol{k}s}+i\eta},
\end{equation}
where $\eta$ is a positive infinitesimal. $J_{\alpha}$
stands for the velocity operator in the eigenvectors space and is given
by $J_{\alpha}=U^{\dagger}j_{\alpha}U$, where $j_{\alpha}=ev_{\alpha}$
is the current operator of electron in the spin space and $v_{\alpha}=\partial H_{\mathrm{mid}}/\partial\left(\hbar k_{\alpha}\right)$
refers to the velocity operator with $\alpha=x,y$. The spin current operators
are represented by
\begin{eqnarray}
J_{\alpha}^{\sigma_{i}} & = & \frac{\hbar}{4}\left\{ v_{\alpha},\sigma_{i}\right\} ,\label{SpinCurrent}
\end{eqnarray}
where $\left\{ A,B\right\} \equiv AB+BA$ is an anti-commutator, and the
$2\times2$ unitary transformation matrix is of the form 
\begin{alignat}{1}
U=\frac{1}{\sqrt{2}}\left(\begin{array}{cc}
-i\varsigma_{\boldsymbol{k}}e^{-i\theta_{\boldsymbol{k}}} & i\varsigma_{\boldsymbol{k}}e^{-i\theta_{\boldsymbol{k}}}\\
1 & 1
\end{array}\right) \;.
\end{alignat}
After taking the trace over the spin degree of freedom, we have the nonzero
component of the intrinsic SHC as
\begin{alignat}{1}
\sigma_{yx}^{\sigma_{z}} & =\frac{eb\lambda\hbar^{2}}{16\pi^{2}}\int_{0}^{k_{c}}\frac{k^{4}dk}{E_{F}-{\hbar^{2}k^{2}}/{2m}}\int_{0}^{2{\pi}}\varsigma_{\boldsymbol{k}}\sin^{2}\theta_{\boldsymbol{k}}\nonumber \\
 & \times\cos2\theta_{\boldsymbol{k}}[\delta(E_{F}-\varepsilon_{\boldsymbol{k}-})-\delta(E_{F}-\varepsilon_{\boldsymbol{k}+})]d\theta_{\boldsymbol{k}},\label{sigmaxyz1}
\end{alignat}
which indicates that a spin Hall current along the $y$ direction
and polarized in the $z$ direction may exist when an external electric
field is applied along the $x$ direction. 
The symbols $b$ and $\lambda$ are defined as
$b={\hbar}/m$ and $\lambda={\beta}{\hbar}^{2}$, respectively. In the weak anisotropy limit [$\beta \ll (2m\hbar k_F)^{-1}$], 
we keep the leading-order contribution to the intrinsic SHC and find
\begin{alignat}{1}
\sigma_{yx}^{\sigma_{z}} & =-\frac{e}{8\pi},
\end{alignat}
which is identical to that of the 2DEG with $k$-linear Rashba SOC
\cite{SinovaPRL} but is different from 2DEG with the $k$-cubed
Rashba SOC~\cite{Schliemann,SHCcubic}. 
The vanishment of the other components of the spin conductivity tensor is due to the symmetry of the Fermi surface.

Now we consider the impact of disorder on the SHC up to the vertex correction. It is more convenient to implement the calculation in the eigenvector space. We consider the randomly distributed, identical point
defects that are spin independent, $V\left(\boldsymbol{r}\right)=\sigma_{0}V_0\sum_{i}\delta\left(\boldsymbol{r}-\boldsymbol{R}_{i}\right)$,
and the matrix element can be expressed as
\begin{align}
V_{\boldsymbol{k}\boldsymbol{k}^{\prime}}^{ss^{\prime}} & =\frac{V_0}{2L^{2}}\sum_{i}e^{-i\left(\boldsymbol{k}-\boldsymbol{k}^{\prime}\right)\cdot\boldsymbol{R}_{i}}\left(1+ss^{\prime}\varsigma_{\boldsymbol{k}}\varsigma_{\boldsymbol{k}^{\prime}}e^{-i\left(\theta_{\boldsymbol{k}^{\prime}}-\theta_{\boldsymbol{k}}\right)}\right),
\end{align}
where $V_0$ is the strength of defect potential and $\boldsymbol{R}_i$ is the position of the defect. The self-energy in the first-order Born approximation can be written as
\begin{align}
 & \left\langle \left\langle \boldsymbol{k}s\right|VG_{0}V\left|\boldsymbol{k}^{\prime}s^{\prime}\right\rangle \right\rangle _{AV}\nonumber \\
= & \frac{nV^{2}}{4L^{2}}\delta_{\boldsymbol{k}\boldsymbol{k}^{\prime}}\sum_{\boldsymbol{k}_{1}s_{1}}g_{\boldsymbol{k}_{1}s_{1}}\left(1+ss^{\prime}\varsigma_{\boldsymbol{k}}\varsigma_{\boldsymbol{k}^{\prime}}\right)\nonumber \\
= & \delta_{s_{1}s_{2}}\delta_{\boldsymbol{k}\boldsymbol{k}^{\prime}}\frac{nV^{2}}{2L^{2}}\sum_{\boldsymbol{k}_{1}s_{1}}g_{\boldsymbol{k}_{1}s_{1}}=\delta_{\boldsymbol{k}\boldsymbol{k}^{\prime}}\delta_{ss^{\prime}}\Sigma_{\boldsymbol{k}s},
\end{align}
where $n=N/L^{2}$ is the density of impurities per unit area and $\left\langle\langle \cdots\right\rangle\rangle _{AV}$ denotes the ensemble
averaging over the impurity distribution. We
have introduced the relation of disorder-free Green's function, $\left\langle \boldsymbol{k}_{1}s_{1}\right|G_{0}\left|\boldsymbol{k}_{2}s_{2}\right\rangle =\delta_{s_{1}s_{2}}\delta_{\boldsymbol{k}_{1}\boldsymbol{k}_{2}}g_{\boldsymbol{k}_{1}s_{1}}$.
Thus the disordered Green's function turns out to be
\begin{equation}
\left\langle \left\langle \boldsymbol{k}s\right|G\left|\boldsymbol{k}^{\prime}s^{\prime}\right\rangle \right\rangle _{AV}=\frac{1}{g_{\boldsymbol{k}s}^{-1}-\Sigma_{\boldsymbol{k}s}}\delta_{ss^{\prime}}\delta_{\boldsymbol{k}^{\prime}\boldsymbol{k}}=\tilde{G}_{\boldsymbol{k}s}.
\end{equation}
For the ladder diagram correction to the velocity operator, we have the following iterative equation
\begin{alignat}{1}
\tilde{v}_{s_1,s_2}^{x}\left(\boldsymbol{k}\right) & =v_{s_1,s_2}^{x}\left(\boldsymbol{k}\right)+\sum_{\boldsymbol{k}^{\prime}}\sum_{s_3,s_4}\left\langle\left\langle V_{\boldsymbol{k}\boldsymbol{k}^{\prime}}^{s_1s_3}V_{\boldsymbol{k}^{\prime}\boldsymbol{k}}^{s_4s_2}\right\rangle\right\rangle _{AV}\nonumber \\
 & \times\tilde{G}_{s_4}^{R}\left(\boldsymbol{k}^{\prime}\right)\tilde{G}_{s_3}^{A}\left(\boldsymbol{k}^{\prime}\right)\tilde{v}_{s_3,s_4}^{x}\left(\boldsymbol{k}^{\prime}\right),
\end{alignat}
where $\tilde{v}^{x}$ is the corrected velocity operator, $s_{1,2,3,4}=\pm1$. 

It is difficult to solve analytically the above self-consistent equation due to the anisotropic dispersion. 
However, by considering the weak SOC limit, i.e., $\mathrm{Im}\Sigma_{k_F} \ll\beta\hbar^3k_F^3\ll\frac{\hbar^2k_F^2}{2m}$, 
the equation can be approximately solved by keeping the leading order of $\beta$, where $k_F$ is the Fermi wave vector 
and $\Sigma_{k_F}$ is the self-energy.

After lengthy but straightforward calculations, we can find the corrected velocity operator
(its derivation is presented in the Supplemental Material \cite{SM1}),
\begin{equation}
\tilde{v}^{x}\left(\boldsymbol{k}\right)=v^{x}\left(\boldsymbol{k}\right)+\beta mE_{F}\sigma_{y}.
\end{equation}
%
Following the similar procedure in Eq.~$\left(\ref{SHC}\right)$, we can calculate the SHC with the vertex correction 
in the weak anisotropy limit and get
\begin{equation}
\left[\sigma_{yx}^{\sigma_{z}}\right]_{\mathrm{V}}=-\frac{e}{16\pi}.
\end{equation}
It can be seen that in the weak anisotropy limit, the vertex correction
reduces the magnitude of SHC by a factor of 2. In fact, 
this unique feature of SHC
under the influence of disorder originates from the special form of
Rashba SOC.
Our result is qualitatively consistent with the fact that the term
 $\alpha k^{2}\left(\boldsymbol{k}\times\boldsymbol{\sigma}\right)\cdot\hat{z}$
would result in a nonzero SHC even with the
vertex correction \cite{MurakamiPRB}. On the other hand,
the vertex correction of disorder can cause the intrinsic SHC of 2DEG with standard $k$-linear 
Rashba SOC  to vanish identically \cite{Inoue}, 
but does not affect the one with $k$-cubed Rashba SOC~\cite{Bernevig}.
Hence, the distinct behaviors of SHC can be used to determine the nature of the
Rashba SOC at LAO/STO interface.
%


In summary, we have developed an effective Hamiltonian of the 2DEG
at the oxide interface LAO/STO and found a different anisotropic Rashba SOC. 
We have found that the static spin susceptibilities are anisotropic and dependent on the Fermi energy. 
We have also demonstrated that this different Rashba SOC possesses entirely different behavior for the SHC 
under disorder. 
Therefore, these unconventional spin responses can be used to determine the nature of Rashba SOC
in experiments.


We are grateful to Matthew Daniels for a careful reading of the manuscript. This work is supported by AFOSR No. FA9550-12-1-0479 and No. FA9550-14-1-0277.

%
%

\onecolumngrid

\section*{Supplementary material for"Spin responses and effective Hamiltonian for the two dimensional
electron gas at oxide interface $\mathrm{LaAlO_{3}/SrTiO_{3}}$"}

In this supplementary material, we provide the detailed derivation of the effective Hamiltonians and the corrected velocity operator.

\section{The derivation of the effective Hamiltonians}

In this section, we construct an effective Hamiltonian around $\Gamma$ point. We first decompose
the TB Hamiltonian $H_{\mathrm{TB}}$ into two parts

\begin{alignat}{1}
H_{\mathrm{TB}} & \approx H\left(k=0\right)+H\left(k\right),
\end{alignat}
where $H\left(k=0\right)$ and $H\left(k\right)$ stand for the Hamiltonian
at the exact $\Gamma$ point and its deviation, respectively. In a new
basis $\left\{ \left|d_{xy}\uparrow\right\rangle ,\left|d_{xz}\downarrow\right\rangle ,\left|d_{yz}\downarrow\right\rangle, \left|d_{xy}\downarrow\right\rangle ,\left|d_{xz}\uparrow\right\rangle ,\left|d_{yz}\uparrow\right\rangle \right\} $,
$H\left(k=0\right)$ and $H\left(k\right)$ take the following form

\begin{alignat}{1}
H\left(k=0\right) & =\left(\begin{array}{cccccc}
-\Delta_{E}-4t & -i\xi & \xi & 0 & 0 & 0\\
i\xi & -2t & i\xi & 0 & 0 & 0\\
\xi & -i\xi & -2t & 0 & 0 & 0\\
0 & 0 & 0 & -\Delta_{E}-4t & -i\xi & -\xi\\
0 & 0 & 0 & i\xi & -2t & -i\xi\\
0 & 0 & 0 & -\xi & i\xi & -2t
\end{array}\right)
\end{alignat}
and

\begin{alignat}{1}
H\left(k\right) & =\left(\begin{array}{cccccc}
tk^{2} & 0 & 0 & 0 & -2i\gamma k_{y} & -2i\gamma k_{x}\\
0 & tk_{x}^{2} & 0 & 2i\gamma k_{y} & 0 & 0\\
0 & 0 & tk_{y}^{2} & 2i\gamma k_{x} & 0 & 0\\
0 & -2i\gamma k_{y} & -2i\gamma k_{x} & tk^{2} & 0 & 0\\
2i\gamma k_{y} & 0 & 0 & 0 & tk_{x}^{2} & 0\\
2i\gamma k_{x} & 0 & 0 & 0 & 0 & tk_{y}^{2}
\end{array}\right).
\end{alignat}
It can be seen that in $H\left(k\right)$ we just keep the leading
order term in $k$ for each component with $k^{2}=k_{x}^{2}+k_{y}^{2}$. At the $\Gamma$
point, we can get the two-fold degenerate eigenvalues

\begin{alignat}{1}
E_{t} & =-\xi-2t\\
E_{m} & =\frac{-\Delta_{E}-6t+\xi}{2}+\sqrt{\left(\frac{\Delta_{E}+2t+\xi}{2}\right)^{2}+2\xi^{2}}\\
E_{b} & =\frac{-\Delta_{E}-6t+\xi}{2}-\sqrt{\left(\frac{\Delta_{E}+2t+\xi}{2}\right)^{2}+2\xi^{2}}
\end{alignat}
and their eigenstates

\begin{alignat}{1}
\psi_{t1} & =C_{1}\left(\theta_{1}\left|d_{xy},\uparrow\right\rangle +i\left|d_{xz},\downarrow\right\rangle +\left|d_{yz},\downarrow\right\rangle \right)\\
\psi_{t2} & =C_{1}\left(-\theta_{1}\left|d_{xy},\downarrow\right\rangle -i\left|d_{xz},\uparrow\right\rangle +\left|d_{yz},\uparrow\right\rangle \right)\\
\psi_{m1} & =\frac{1}{\sqrt{2}}\left(-i\left|d_{xz},\downarrow\right\rangle +\left|d_{yz},\downarrow\right\rangle \right)\\
\psi_{m2} & =\frac{1}{\sqrt{2}}\left(i\left|d_{xz},\uparrow\right\rangle +\left|d_{yz},\uparrow\right\rangle \right)\\
\psi_{b1} & =C_{2}\left(\theta_{2}\left|d_{xy},\uparrow\right\rangle +i\left|d_{xz},\downarrow\right\rangle +\left|d_{yz},\downarrow\right\rangle \right)\\
\psi_{b2} & =C_{2}\left(-\theta_{2}\left|d_{xy},\downarrow\right\rangle -i\left|d_{xz},\uparrow\right\rangle +\left|d_{yz},\uparrow\right\rangle \right),
\end{alignat}
where the parameters are given by

\begin{equation}
\theta_{1} =\frac{E_{t}+2t-\xi}{\xi},\:\theta_{2}=\frac{E_{b}+2t-\xi}{\xi},\: C_{1,2}  =\frac{1}{\sqrt{\theta_{1,2}^{2}+2}}.
\end{equation}
Here the subscripts $t,m,b$ refer to top, middle and bottom bands, respectively. In the basis $\left\{ \psi_{t1},\psi_{t2},\psi_{m1},\psi_{m2},\psi_{b1},\psi_{b2}\right\} $,
the Hamiltonian $H_{\mathrm{TB}}$ can be expressed as

\begin{alignat*}{1}
H_{\mathrm{TB}} & =\left(\begin{array}{cccccc}
t_{11}k^{2} & it_{12}k_{-} & t_{13}\left(k_{x}^{2}-k_{y}^{2}\right) & it_{14}k_{+} & t_{15}k^{2} & it_{16}k_{-}\\
 & t_{11}k^{2} & -it_{14}k_{-} & t_{13}\left(k_{x}^{2}-k_{y}^{2}\right) & -it_{16}k_{+} & t_{15}k^{2}\\
 &  & tk^{2}/2 & 0 & t_{35}\left(k_{x}^{2}-k_{y}^{2}\right) & it_{36}k_{+}\\
 &  &  & tk^{2}/2 & -it_{36}k_{-} & t_{35}\left(k_{x}^{2}-k_{y}^{2}\right)\\
 & * &  &  & t_{55}k^{2} & it_{56}k_{-}\\
 &  &  &  &  & t_{55}k^{2}
\end{array}\right),
\end{alignat*}
where the parameters are $t_{11}=C_{1}^{2}\left(1+\theta_{1}^{2}\right)t$,
$t_{12}=-4C_{1}^{2}\gamma\theta_{1}$, $t_{13}=-C_{1}t/\sqrt{2}$,
$t_{56}=-4C_{2}^{2}\gamma\theta_{2}$, $t_{35}=-C_{2}t/\sqrt{2}$,
$t_{14}=-2C_{1}\theta_{1}\gamma/\sqrt{2}$, $t_{15}=C_{1}C_{2}\left(1+\theta_{1}\theta_{2}\right)t$,
$t_{16}=-2\gamma C_{1}C_{2}\left(\theta_{1}+\theta_{2}\right)$, $t_{36}=-2C_{2}\theta_{2}\gamma/\sqrt{2}$
and $t_{55}=C_{2}^{2}\left(1+\theta_{2}^{2}\right)t$.

We follow the standard quasi-degenerate perturbation theory and seperate the Hamiltonian into the unperturbated and perturbated
part $H_{\mathrm{TB}}=H_{\mathrm{np}}+H_{\mathrm{per}}$. The corresponding matrix elements of perturbation Hamiltonian $H_{\mathrm{per}}$ read

\begin{alignat}{1}
\left\langle \psi_{t1}\right|H_{\mathrm{per}}\left|\psi_{t1}\right\rangle  & =\frac{t_{13}^{2}\left(k_{x}^{2}-k_{y}^{2}\right)^{2}+t_{14}^{2}k^{2}}{E_{m}-E_{t}}+\frac{t_{15}^{2}k^{4}+t_{16}^{2}k^{2}}{E_{b}-E_{t}}\\
\left\langle \psi_{t1}\right|H_{\mathrm{per}}\left|\psi_{t2}\right\rangle  & =\frac{2it_{13}t_{14}\left(k_{x}^{2}-k_{y}^{2}\right)k_{+}+t_{14}^{2}k^{2}}{E_{m}-E_{t}}+\frac{2it_{15}t_{16}k^{2}k_{-}}{E_{b}-E_{t}}\\
\left\langle \psi_{m1}\right|H_{\mathrm{per}}\left|\psi_{m1}\right\rangle  & =\frac{t_{13}^{2}\left(k_{x}^{2}-k_{y}^{2}\right)^{2}+t_{14}^{2}k^{2}}{E_{t}-E_{m}}+\frac{t_{35}^{2}\left(k_{x}^{2}-k_{y}^{2}\right)^{2}+t_{36}^{2}k^{2}}{E_{b}-E_{m}}\\
\left\langle \psi_{m1}\right|H_{\mathrm{per}}\left|\psi_{m2}\right\rangle  & =2i\left(\frac{t_{13}t_{14}}{E_{t}-E_{m}}+\frac{t_{35}t_{36}}{E_{b}-E_{m}}\right)\left(k_{x}^{2}-k_{y}^{2}\right)k_{+}\\
\left\langle \psi_{b1}\right|H_{\mathrm{per}}\left|\psi_{b1}\right\rangle  & =\frac{t_{13}^{2}\left(k_{x}^{2}-k_{y}^{2}\right)^{2}+t_{14}^{2}k^{2}}{E_{t}-E_{b}}+\frac{t_{35}^{2}k^{4}+t_{36}^{2}k^{2}}{E_{m}-E_{b}}\\
\left\langle \psi_{b1}\right|H_{\mathrm{per}}\left|\psi_{b2}\right\rangle  & =\frac{2it_{15}t_{16}k^{2}k_{-}}{E_{t}-E_{b}}+\frac{2it_{35}t_{36}\left(k_{x}^{2}-k_{y}^{2}\right)k_{+}}{E_{m}-E_{b}}
\end{alignat}
and the others can be obtained by the relations

\begin{alignat}{1}
\left\langle \psi_{a1}\right|H_{\mathrm{per}}\left|\psi_{a1}\right\rangle  & =\left\langle \psi_{a2}\right|H_{\mathrm{per}}\left|\psi_{a2}\right\rangle \\
\left\langle \psi_{a2}\right|H_{\mathrm{per}}\left|\psi_{a1}\right\rangle ^{\star} & =\left\langle \psi_{a1}\right|H_{\mathrm{per}}\left|\psi_{a2}\right\rangle,
\end{alignat}
where $a=t,m,b$ are the band indices. Up to the leading order of
the Rashba SOC, we could get the effective Hamiltonian for the top
pair of bands

\begin{alignat}{1}
H_{t}\left(\boldsymbol{k}\right) & =\left(\begin{array}{cc}
\frac{k^{2}}{2m_{t}} & i\alpha_{t}k_{-}\\
-i\alpha_{t}k_{+} & \frac{k^{2}}{2m_{t}}
\end{array}\right),
\end{alignat}
for the middle pair of bands

\begin{alignat}{1}
H_{m}\left(\boldsymbol{k}\right) & =\left(\begin{array}{cc}
\frac{k^{2}}{2m_{m}} & i\alpha_{m}\left(k_{x}^{2}-k_{y}^{2}\right)k_{+}\\
-i\alpha_{m}\left(k_{x}^{2}-k_{y}^{2}\right)k_{-} & \frac{k^{2}}{2m_{m}}
\end{array}\right),
\end{alignat}
and for the bottom pair of bands

$ $
\begin{alignat}{1}
H_{b}\left(\boldsymbol{k}\right) & =\left(\begin{array}{cc}
\frac{k^{2}}{2m_{b}} & i\alpha_{b}k_{-}\\
-i\alpha_{b}k_{+} & \frac{k^{2}}{2m_{b}}
\end{array}\right),
\end{alignat}
where the effective masses $m_{a}$ and Rashba SOC strength $\alpha_{a}$
($a=t,m,b$) are given by

\begin{alignat}{1}
\frac{1}{2m_{t}} & =t_{11}+\frac{t_{14}^{2}}{E_{m}-E_{t}}+\frac{t_{16}^{2}}{E_{b}-E_{t}}\\
\frac{1}{2m_{m}} & =\frac{t}{2}+\frac{t_{14}^{2}}{E_{t}-E_{m}}+\frac{t_{36}^{2}}{E_{b}-E_{m}}\\
\frac{1}{2m_{b}} & =t_{55}+\frac{t_{16}^{2}}{E_{t}-E_{b}}+\frac{t_{36}^{2}}{E_{m}-E_{b}}
\end{alignat}

\begin{alignat}{1}
\alpha_{t} & =t_{12},\:\alpha_{b}=t_{56}\\
\alpha_{m} & =2i\left(\frac{t_{13}t_{14}}{E_{t}-E_{m}}+\frac{t_{35}t_{36}}{E_{b}-E_{m}}\right).
\end{alignat}

\section{The derivation of the corrected velocity operator}

In this section, we turn to give a detailed calculation of vertex correction. As shown in the main text, we have the iterative equation
\begin{alignat}{1}\label{iterative0}
\tilde{v}_{s_1,s_2}^{x}\left(\boldsymbol{k}\right) & =v_{s_1,s_2}^{x}\left(\boldsymbol{k}\right)+\sum_{\boldsymbol{k}^{\prime}}\sum_{s_3,s_4}\left\langle\left\langle V_{\boldsymbol{k}\boldsymbol{k}^{\prime}}^{s_1s_3}V_{\boldsymbol{k}^{\prime}\boldsymbol{k}}^{s_4s_2}\right\rangle\right\rangle _{AV}\nonumber \\
 & \times\tilde{G}_{s_4}^{R}\left(\boldsymbol{k}^{\prime}\right)\tilde{G}_{s_3}^{A}\left(\boldsymbol{k}^{\prime}\right)\tilde{v}_{s_3,s_4}^{x}\left(\boldsymbol{k}^{\prime}\right),
\end{alignat}
where Green's function $\tilde{G}_s^{R/A}=1/(E_F-\epsilon_{\boldsymbol{k}s}\mp i\mathrm{Im}\Sigma_{\boldsymbol{k}s})$ with $\mathrm{Im}\Sigma_{\boldsymbol{k}s}\equiv\hbar/2\tau_{\boldsymbol{k}s}$ and $s_{1,2,3,4}=\pm1$. Eigenvalue reads
\begin{align}
\epsilon_{\boldsymbol{k}s}&=\frac{\hbar^2k^2}{2m}+s\beta\hbar^3k^3|\cos2\theta_{\boldsymbol{k}}|,
\end{align}
and wave function reads
\begin{align}
\phi_{\boldsymbol{k}s}&=\frac{e^{i\boldsymbol{k}\cdot\boldsymbol{r}}}{\sqrt{2}L}\left(\begin{array}{cc}
-is\zeta_{\boldsymbol{k}}e^{-i\theta_{\boldsymbol{k}}} \\
1 \\
\end{array}\right),\    \ \zeta_{\boldsymbol{k}}=\cos2\theta_{\boldsymbol{k}}/|\cos2\theta_{\boldsymbol{k}}|.
\end{align}
Velocity matrix element in the eigenvector space reads
\begin{align}\label{velocity_matrix}
\left(\begin{array}{cc}
(v^x_{\boldsymbol{k}})_{++} & (v^x_{\boldsymbol{k}})_{+-} \\
(v^x_{\boldsymbol{k}})_{-+} & (v^x_{\boldsymbol{k}})_{--} \\
\end{array}\right)&=\left(\begin{array}{cc}
\frac{\hbar k}{m}\cos\theta_{\boldsymbol{k}}+\frac{\beta\hbar^2k^2}{2}\zeta_{\boldsymbol{k}}(\cos3\theta_{\boldsymbol{k}}+5\cos\theta_{\boldsymbol{k}}) & \frac{i\beta\hbar^2k^2}{2}\zeta_{\boldsymbol{k}}(\sin3\theta_{\boldsymbol{k}}-\sin\theta_{\boldsymbol{k}}) \\
-\frac{i\beta\hbar^2k^2}{2}\zeta_{\boldsymbol{k}}(\sin3\theta_{\boldsymbol{k}}-\sin\theta_{\boldsymbol{k}}) & \frac{\hbar k}{m}\cos\theta_{\boldsymbol{k}}-\frac{\beta\hbar^2k^2}{2}\zeta_{\boldsymbol{k}}(\cos3\theta_{\boldsymbol{k}}+5\cos\theta_{\boldsymbol{k}}) \\
\end{array}\right),
\end{align}
where $(v^x_{\boldsymbol{k}})_{ss^{\prime}}\equiv\langle\phi_{\boldsymbol{k}s}|v^x|\phi_{\boldsymbol{k}s^{\prime}}\rangle$. Based on these, one can evaluate the disorder-averaged correlation function
\begin{align}
\left\langle\left\langle V^{++}_{\boldsymbol{k},\boldsymbol{k}^{'}}V^{++}_{\boldsymbol{k}^{'},\boldsymbol{k}}\right\rangle\right\rangle _{AV}
&=\left\langle\left\langle V^{--}_{\boldsymbol{k},\boldsymbol{k}^{'}}V^{--}_{\boldsymbol{k}^{'},\boldsymbol{k}}\right\rangle\right\rangle _{AV}
=\left\langle\left\langle V^{--}_{\boldsymbol{k}^{'},\boldsymbol{k}}V^{++}_{\boldsymbol{k},\boldsymbol{k}^{'}}\right\rangle\right\rangle _{AV}
=\frac{nV_0^2}{2L^2}(1+\zeta_{\boldsymbol{k}}\zeta_{\boldsymbol{k}^{'}}\cos(\theta_{\boldsymbol{k}}-\theta_{\boldsymbol{k}^{'}})),\\
\left\langle\left\langle V^{+-}_{\boldsymbol{k},\boldsymbol{k}^{'}}V^{-+}_{\boldsymbol{k}^{'},\boldsymbol{k}}\right\rangle\right\rangle _{AV}
&=\left\langle\left\langle V^{-+}_{\boldsymbol{k},\boldsymbol{k}^{'}}V^{+-}_{\boldsymbol{k}^{'},\boldsymbol{k}}\right\rangle\right\rangle _{AV}
=\left\langle\left\langle V^{+-}_{\boldsymbol{k}^{'},\boldsymbol{k}}V^{+-}_{\boldsymbol{k},\boldsymbol{k}^{'}}\right\rangle\right\rangle _{AV}
=\frac{nV_0^2}{2L^2}(1-\zeta_{\boldsymbol{k}}\zeta_{\boldsymbol{k}^{'}}\cos(\theta_{\boldsymbol{k}}-\theta_{\boldsymbol{k}^{'}})),\\
\left\langle\left\langle V^{-+}_{\boldsymbol{k}^{'},\boldsymbol{k}}V^{++}_{\boldsymbol{k},\boldsymbol{k}^{'}}\right\rangle\right\rangle _{AV}
&=-\left\langle\left\langle V^{++}_{\boldsymbol{k}^{'},\boldsymbol{k}}V^{+-}_{\boldsymbol{k},\boldsymbol{k}^{'}}\right\rangle\right\rangle _{AV}
=-\left\langle\left\langle V^{--}_{\boldsymbol{k}^{'},\boldsymbol{k}}V^{+-}_{\boldsymbol{k},\boldsymbol{k}^{'}}\right\rangle\right\rangle _{AV}
=\frac{nV_0^2}{2L^2}i\zeta_{\boldsymbol{k}}\zeta_{\boldsymbol{k}^{'}}\sin(\theta_{\boldsymbol{k}}-\theta_{\boldsymbol{k}^{'}}).
\end{align}
Now we focus on the weak scattering and weak anisotropy limit, i.e., $\mathrm{Im}\Sigma_{k_F}\ll\beta\hbar^3k_F^3\ll\frac{\hbar^2k_F^2}{2m}$. In this sense, 
we have the approximated expression of Fermi wave vector and density of states
\begin{align}
k_{s,F}&\approx\frac{\sqrt{2mE_F}}{\hbar}-s\frac{2\beta m^2E_F}{\hbar}|\cos2\theta_{\boldsymbol{k}}|,\\
N_{s,F}&\approx\frac{m}{2\pi\hbar^2}-s\frac{3\beta m^2}{\pi^2\hbar^2}\sqrt{2mE_F},
\end{align}
where $k_{\pm,F}$ and $N_{\pm,F}$ are the Fermi wave vector and density of states for two branches of bands. And the relaxation time $\tau_{\boldsymbol{k}s}$ is given by
\begin{align}\nonumber
\frac{1}{\tau_{\boldsymbol{k},+}}&=\frac{1}{\tau_{\boldsymbol{k},-}}=\frac{2\pi}{\hbar}\sum_{\boldsymbol{k}^{'}}\left\langle\left\langle V_{\boldsymbol{k},\boldsymbol{k}^{'}}^{++}V_{\boldsymbol{k}^{'},\boldsymbol{k}}^{++}\right\rangle\right\rangle _{AV}\delta(E_F-\epsilon_{\boldsymbol{k}^{'},+})+\frac{2\pi}{\hbar}\sum_{\boldsymbol{k}^{'}}\left\langle\left\langle V_{\boldsymbol{k},\boldsymbol{k}^{'}}^{+-}V_{\boldsymbol{k}^{'},\boldsymbol{k}}^{-+}\right\rangle\right\rangle _{AV}\delta(E_F-\epsilon_{\boldsymbol{k}^{'},-})\\
&=\frac{nV_0^2m}{\hbar^3}.
\end{align}
According to Eq. (\ref{velocity_matrix}), it is natural to assume that the modified velocity takes the following form
\begin{eqnarray}
(\tilde{v}^x_{\boldsymbol{k}})_{++}&=&(\beta A_0\zeta_{\boldsymbol{k}}+A_{1}k)\cos\theta_{\boldsymbol{k}}+\beta k^2\zeta_{\boldsymbol{k}}(A_2\cos3\theta_{\boldsymbol{k}}+A_3\cos\theta_{\boldsymbol{k}}),\\
(\tilde{v}^x_{\boldsymbol{k}})_{+-}&=&\beta B_0\zeta_{\boldsymbol{k}}\sin\theta_{\boldsymbol{k}}+\beta k^2\zeta_{\boldsymbol{k}}(B_1\sin3\theta_{\boldsymbol{k}}+B_2\sin\theta_{\boldsymbol{k}}),\\
(\tilde{v}^x_{\boldsymbol{k}})_{-+}&=&\beta C_0\zeta_{\boldsymbol{k}}\sin\theta_{\boldsymbol{k}}+\beta k^2\zeta_{\boldsymbol{k}}(C_1\sin3\theta_{\boldsymbol{k}}+C_2\sin\theta_{\boldsymbol{k}}),\\
(\tilde{v}^x_{\boldsymbol{k}})_{--}&=&(\beta D_0\zeta_{\boldsymbol{k}}+D_{1}k)\cos\theta_{\boldsymbol{k}}+\beta k^2\zeta_{\boldsymbol{k}}(D_2\cos3\theta_{\boldsymbol{k}}+D_3\cos\theta_{\boldsymbol{k}}).
\end{eqnarray}
Substitute these matrix elements into Eq. (\ref{iterative0}), and by using the relations
\begin{eqnarray}\nonumber
\int\frac{kdk}{2\pi}\tilde{G}^R_{\boldsymbol{k},+}\tilde{G}^A_{\boldsymbol{k},-}
&=&\int\frac{d\epsilon_{\boldsymbol{k},-}}{2(\frac{\hbar^2}{m}-3\beta\hbar^3 k|\cos2\theta_{\boldsymbol{k}}|)}\frac{i\delta(E_F-\epsilon_{\boldsymbol{k},-})}{E_F-\epsilon_{\boldsymbol{k},-}-2\beta\hbar^3k^3|\cos2\theta_{\boldsymbol{k}}|}\\\nonumber
&-&\int\frac{d\epsilon_{\boldsymbol{k},+}}{2(\frac{\hbar^2}{m}+3\beta\hbar^3k|\cos2\theta_{\boldsymbol{k}}|)}
\frac{i\delta(E_F-\epsilon_{\boldsymbol{k},+})}{E_F-\epsilon_{\boldsymbol{k},+}+2\beta\hbar^3 k^3|\cos2\theta_{\boldsymbol{k}}|}\\
&=&-\frac{m}{\hbar^2}\frac{i}{2\beta\hbar^3k_F^3|\cos2\theta_{\boldsymbol{k}}|},\\
\int\frac{kdk}{2\pi}\tilde{G}^R_{\boldsymbol{k},-}\tilde{G}^A_{\boldsymbol{k},+}&=&\frac{m}{\hbar^2}\frac{i}{2\beta\hbar^3 k_F^3|\cos2\theta_{\boldsymbol{k}}|},
\end{eqnarray}
one finally obtains these coefficients of $A_{i}$,~$B_{i}$,~$C_{i}$ and $D_{i}$, 
\begin{eqnarray}\nonumber
&& A_0=mE_F,\   \  A_1=\frac{\hbar}{m},\    \  A_2=\frac{\hbar^2}{2},\    \  A_3=\frac{5\hbar^2}{2},\\\nonumber
&& B_0=imE_F,\    \  B_1=\frac{i\hbar^2}{2},\    \ B_2=-\frac{i\hbar^2}{2},\\\nonumber
&& C_0=-imE_F,\    \  C_1=-\frac{i\hbar^2}{2},\    \ C_2=\frac{i\hbar^2}{2},\\
&& D_0=-mE_F,\   \  D_1=\frac{\hbar}{m},\    \  D_2=-\frac{\hbar^2}{2},\    \  D_3=-\frac{5\hbar^2}{2}.
\end{eqnarray}
This indicates a relation between the unmodified and modified velocities (in eigenvector space)
\begin{align}
\left(\begin{array}{cc}
(\tilde{v}^x_{\boldsymbol{k}})_{++} & (\tilde{v}^x_{\boldsymbol{k}})_{+-} \\
(\tilde{v}^x_{\boldsymbol{k}})_{-+} & (\tilde{v}^x_{\boldsymbol{k}})_{--} \\
\end{array}\right)&=\left(\begin{array}{cc}
(v^x_{\boldsymbol{k}})_{++} & (v^x_{\boldsymbol{k}})_{+-} \\
(v^x_{\boldsymbol{k}})_{-+} & (v^x_{\boldsymbol{k}})_{--} \\
\end{array}\right)+\beta mE_F\zeta_{\boldsymbol{k}}\left(\begin{array}{cc}
\cos\theta_{\boldsymbol{k}} & i\sin\theta_{\boldsymbol{k}} \\
-i\sin\theta_{\boldsymbol{k}} & -\cos\theta_{\boldsymbol{k}} \\
\end{array}\right).
\end{align}
When transformed into the spin space, this means
\begin{align}
\tilde{v}^x_{\boldsymbol{k}}&=v^x_{\boldsymbol{k}}+\beta mE_F\sigma_y,
\end{align}
which is nothing but the corrected velocity operator in Eq.~(26) in the main text.


\begin{thebibliography}{References}
\bibitem{touching} The two branches of anisotropic Rashba SOC do not actually touch each other when a higher-order correction is taken into account.
\bibitem{SUSCR} The spin susceptibilities of 2DEG with $k$-cubed Rashba SOC
have the same qualitative features as 2DEG with an anisotropic Rashba SOC but with some quantitative
differences.
\end{thebibliography}
\end{document}